\documentclass[aps,prx,twocolumn,superscriptaddress,floatfix]{revtex4-2}

\usepackage[T1]{fontenc}
\usepackage{microtype}
\usepackage{xspace}
\usepackage{soul}

\usepackage{chemformula}
\usepackage{amsmath}
\usepackage{amssymb}
\usepackage{mathtools}
\usepackage{amsthm}
\usepackage{bm}
\usepackage{cancel}

\usepackage{graphicx}
\usepackage{adjustbox}
\usepackage{subcaption}
\usepackage{float}
\usepackage{flafter}
\usepackage{placeins}

\usepackage{algorithm}
\usepackage{algpseudocode}
\algdef{SE}[SUBALG]{Indent}{EndIndent}{}{\algorithmicend\ }%
\algtext*{Indent}
\algtext*{EndIndent}

\usepackage{booktabs}
\usepackage{multirow}
\usepackage{threeparttable}
\usepackage{tablefootnote}
\usepackage{caption}

\usepackage{url}

\usepackage{xcolor,colortbl}
\usepackage[normalem]{ulem}
\usepackage{xparse}

\usepackage{titlesec}
\titleformat{\section}[block]{\normalfont\bfseries}{}{0em}{}
\usepackage[hidelinks]{hyperref}
\usepackage[capitalize,noabbrev]{cleveref}
\usepackage{xr}
\usepackage{caption}
\theoremstyle{plain}

\theoremstyle{definition}

\theoremstyle{remark}

\usepackage{enumitem}
\usepackage{eucal}

\usepackage{lineno}
\ifdefined\linenomathpatchAMS\linenomathpatchAMS\fi

\begin{document}

\title{Distilling Large-Scale Machine Learning Interatomic Potentials to Predict Long-Timescale Electrolyte Transport Properties}

\author{Ishan Amin}
\thanks{These authors contributed equally to this work.}
\affiliation{Department of Computer Science, University of California, Berkeley, Berkeley, California 94720, USA}

\author{Yue Jian}
\thanks{These authors contributed equally to this work.}
\affiliation{Department of Chemical and Biomolecular Engineering, University of California, Berkeley, Berkeley, California 94720, USA}

\author{Nitesh Kumar}
\affiliation{Materials Sciences Division, Lawrence Berkeley National Laboratory, Berkeley, California 94720, USA}

\author{Christine Zhang}
\affiliation{Department of Computer Science, University of California, Berkeley, Berkeley, California 94720, USA}

\author{Jianwei Lai}
\affiliation{Lawrence Berkeley National Laboratory, Berkeley, California 94720, USA}

\author{Casey S. Mezerkor}
\affiliation{Physical and Computational Sciences Directorate, Pacific Northwest National Laboratory, Richland, Washington 99354, USA}

\author{Aaron M. Melemed}
\affiliation{Department of Mechanical Engineering, University of Michigan, Ann Arbor, Michigan 48109, USA}

\author{Jonathan R. Thurston}
\affiliation{Department of Mechanical Engineering, University of Michigan, Ann Arbor, Michigan 48109, USA}

\author{Neil P. Dasgupta}
\affiliation{Department of Mechanical Engineering, University of Michigan, Ann Arbor, Michigan 48109, USA}

\author{J. David Bazak}
\affiliation{Physical and Computational Sciences Directorate, Pacific Northwest National Laboratory, Richland, Washington 99354, USA}

\author{Ethan J. Crumlin}
\affiliation{Lawrence Berkeley National Laboratory, Berkeley, California 94720, USA}

\author{Samuel Blau}
\email{smblau@lbl.gov}
\affiliation{Lawrence Berkeley National Laboratory, Berkeley, California 94720, USA}

\author{Aditi S. Krishnapriyan}
\email{aditik1@berkeley.edu}
\affiliation{Department of Computer Science, University of California, Berkeley, Berkeley, California 94720, USA}
\affiliation{Department of Chemical and Biomolecular Engineering, University of California, Berkeley, Berkeley, California 94720, USA}

\begin{abstract}
Accurately predicting the transport properties of liquid electrolytes is critical for next-generation battery design and requires force fields that are both quantum-mechanically accurate and fast enough for long-timescale simulation. Recent large-scale pretrained machine learning interatomic potentials (MLIPs) approach density-functional-theory accuracy, but remain too slow to access the necessary timescales to converge transport properties. We provide a knowledge distillation procedure to transfer the accuracy of a large-scale "teacher" MLIP into fast, system-specific "students", without any additional ab initio data. Our distilled models run ${\sim}40\times$ faster, reproduce teacher-predicted structural observables, and produce stable 20~ns trajectories in days, rather than the months the teacher would require. Across a diverse set of sodium- and lithium-ion electrolytes, student-predicted self-diffusivities, ionic conductivities, and shear viscosities track experiment and outperform classical force-field predictions, establishing distillation as a practical route to accurate, long-timescale simulations necessary for high-throughput electrolyte screening.

\end{abstract}

\maketitle

\begin{figure*}[t]
    \centering
    \includegraphics[width=\textwidth]{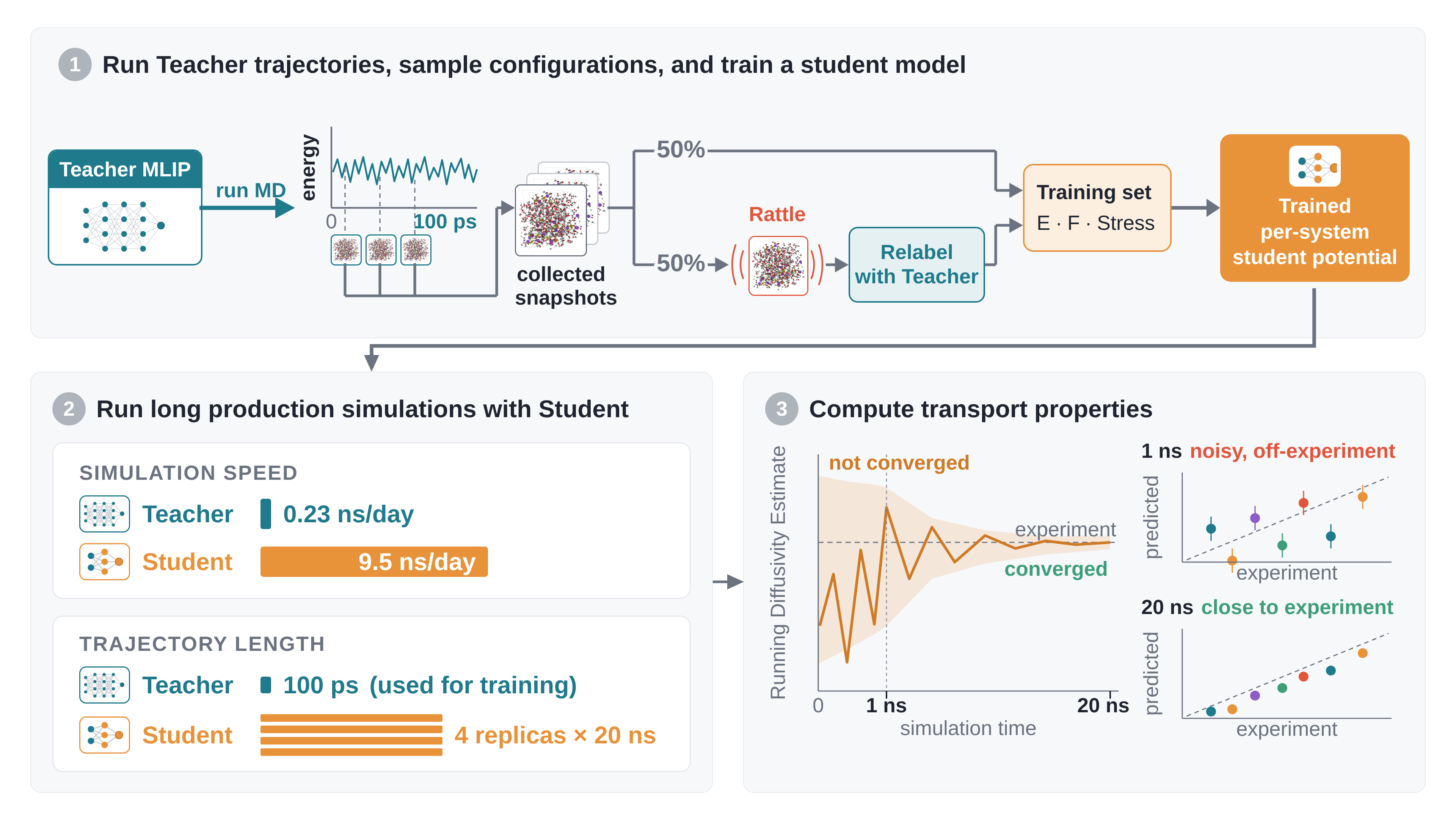}
    \caption{\textbf{Teacher--student distillation workflow for converged electrolyte transport properties.} The workflow distills an accurate but slow teacher potential into a fast, system-specific student that can be run long enough to converge transport coefficients. \textbf{(1)~Data generation and distillation.} The teacher potential first runs a short reference trajectory, from which training configurations are sampled. To broaden coverage beyond near-equilibrium structures, half of the sampled configurations are rattled with Gaussian noise and relabeled with the teacher to provide consistent energy, force, and stress targets. The student potential is then trained on this mixed dataset of both perturbed and unperturbed samples. \textbf{(2)~Production simulations.} Once trained, the student replaces the teacher for production molecular dynamics, running ${\sim}40\times$ faster and reaching trajectory lengths of $20$~ns. \textbf{(3)~Convergence and validation.} The 20~ns trajectories allow transport observables to converge. Transport coefficients start out noisy at short timescales (1~ns), but converge near experimental values once the trajectory is sufficiently long.}
    \label{fig:workflow}
\end{figure*}

Ion transport is an essential factor in battery electrolyte performance. Transport coefficients such as ionic conductivity, the per-species self-diffusivities, and the shear viscosity determine an electrochemical cell's rate capability, power density, and low-temperature effectiveness \cite{xu2004}. Since the number of viable electrolytes is combinatorially large, thorough experimental exploration of the space is costly and time-consuming. Atomistic simulations offer an alternative for rapid screening without requiring manual synthesis and testing, provided the simulation method can both accurately and quickly predict the transport properties of the electrolyte \cite{zhu2024,lai2025,hu2022}.

However, calculating converged transport coefficients through simulation requires not only quantum-mechanically accurate forces and stresses, but also multiple nanoseconds of well-sampled molecular dynamics (MD) simulations averaged over multiple replicas \cite{maginn2019,zhang2015viscosity,ringsby2021}. Evaluating self-diffusivity and ionic conductivity requires calculating the mean-squared displacement across extended timescales, while viscosity relies on time-integrating autocorrelation functions until they fully decay \cite{maginn2019,fong2020}. As a result, these calculations depend on extended lag times, decreasing the number of independent sampling windows, which in turn increases the statistical noise \cite{jones2012}. 
Ab initio methods are too expensive for the millions of steps needed to converge transport properties, while classical force fields, though computationally efficient, often misrepresent solvation structures and ion--ion correlations \cite{borodin_applep}. Per-system fitting to ab initio or experimental data can improve classical force-field accuracy, but is impractical for high-throughput screening.

Machine learning interatomic potentials (MLIPs), which approximate ab initio accuracy at a fraction of the cost \cite{behler2007,schran2021,mace2022,nequip2022,qu2026a,jacobs2025, qu2024importance,behler2026strategic}, have been increasingly deployed to compute electrolyte transport properties, though existing efforts remain limited in chemical scope \cite{bamboo,phyneo_electrolyte,ju2025,byteff_pol,wang_natcomm2025}. Prior works mostly rely on bespoke datasets that would need substantial expansion to cover the entire space of potential battery candidates. BAMBOO and PhyNEO-Electrolyte are trained on data covering at most a few dozen solvents and salts, and validate transport only for lithium electrolytes \cite{bamboo,phyneo_electrolyte}. Ju et al.\ applied SevenNet-0, pretrained on inorganic crystals, zero-shot to lithium electrolytes, but had to fix the simulation box to experimental densities in order to reproduce experimental diffusivities \cite{ju2025}. ByteFF-Pol trains a graph neural network to predict the parameters of a polarizable classical force field; while its dataset includes hundreds of solvent molecules, a significant advancement, it contains only three anions and lithium as the only cation \cite{byteff_pol}. Wang et al.\ go furthest toward a universal electrolyte potential, creating a dataset covering multiple cations, but their dataset is still only 160,000 structures and the model is again only validated on lithium systems \cite{wang_natcomm2025}. In each case, moving beyond the demonstrated scope would require costly large-scale ab initio data generation.

Recently, large-scale MLIPs pretrained on chemically diverse datasets have demonstrated the ability to extrapolate to systems not seen during training, reducing the need for system-specific fitting \cite{mace_mp0,orb,mace_polar,opoly26,yuan2026foundation}. In particular, the Open Molecules 2025 (OMol25) dataset \cite{omol25} thoroughly samples electrolyte configurations across the periodic table, unlocking the potential of models pretrained on the data to accurately simulate diverse electrolyte compositions. Prior work has shown that OMol25-trained MLIPs, such as the Universal Model for Atoms (UMA) \cite{uma}, can closely reproduce short timescale structural properties of sodium-based electrolytes \cite{kumar2026}. However, the size and speed of the models needed to fit such datasets make them too slow to reach the timescales required for converged transport coefficients. 

Knowledge distillation of MLIPs offers a way to use OMol25-pretrained MLIPs to predict electrolyte transport across chemistries, by accessing the speedups granted by smaller models without sacrificing accuracy. A large pretrained MLIP (the ``teacher'') labels configurations of a specific system or region of chemical space through a systematic procedure. Those configurations are used to train a smaller, faster model with fewer parameters (the ``student'') \cite{hinton2015,amin2025,lightpfp,sevennet_nano,wang2026latent}. The student can then simulate in the teacher's place, ideally with the same accuracy as the teacher but orders of magnitude faster, without any new expensive ab initio labels.

The ability of distilled models to accurately predict long-timescale transport behavior, however, has not been systematically tested. Prior distillation work has primarily focused on static observables, such as energies, forces, and radial distribution functions, mostly computed from trajectories of a nanosecond or less \cite{kim2026density,gardner2025distillation, matin2025teacher}. To date, transport properties reported for distilled potentials have largely been limited to a single transport coefficient for a handful of systems per study \cite{dpa2,pfd,zhang2026npj,sevennet_nano,lightpfp,matsumura2025,chen2026distilling,lian2026}. Excellent performance on static observables over short trajectories does not necessarily imply correct dynamical properties, or student fidelity across long timescales \cite{fu2023,stocker2022,raja2025stable, miret2025energyforceregression}. Accurate transport properties therefore require a distilled model to both remain stable and track ion--ion correlations across long, thoroughly sampled simulations.

In this study, we provide a procedure to distill large-scale MLIPs into students capable of running multi-nanosecond simulations, which are necessary to converge transport coefficients. Our simulations both preserve teacher-model accuracy and track experimentally measured transport properties. To our knowledge, this is the first application of knowledge distillation to transport coefficients in liquid electrolytes, validated against experiment across more than a dozen systems. Our distilled models run ${\sim}40\times$ faster than their teacher, and enable simulations two hundred times the length of the trajectories used to generate their training data. This allows us to access dynamical regimes otherwise inaccessible to the original teacher model. To verify that our students are faithful to the teacher, we periodically spot-check the student's energy, force, and stress errors against the teacher with single-point calculations throughout the simulation, at intervals that make it relatively cheap to do so. We apply our approach to sixteen next-generation candidates for battery electrolytes, focusing on sodium-ion systems \cite{hwang2017,yabuuchi2014}, with various solutes, solvents, and concentrations (detailed in the Results).

For each salt--solvent pair, we distill the small variant of UMA evaluated with its OMol task head (checkpoint \texttt{uma-s-1p1}; hereafter referred to as UMA-OMol) into a system-specific student MLIP built on PaiNN, an equivariant graph neural network~\cite{painn}. The student is trained on a set of configurations sampled from a 100~ps teacher trajectory, with some of the structures noised and relabeled to ensure student robustness. We then run four independent 20~ns replica simulations per system, and periodically check the student model's errors against single-point predictions from UMA-OMol along each trajectory to confirm that they remain low and stable. The student trajectories are then used to compute the self-diffusivities, ionic conductivities, and shear viscosities for every electrolyte system, which are compared against experiment. An overview of the workflow is shown in Fig.~\ref{fig:workflow}.

Across all sixteen systems, the distilled potentials capture experimental trends and agree more closely with experiment than the widely used OPLS-AA force field \cite{jorgensen1996}, reducing the median deviation from experiment for every transport property studied. By transferring the accuracy of pretrained MLIPs into compact models, distillation enables high-throughput electrolyte transport simulations that achieve a balance of accuracy and efficiency unavailable from classical force fields, direct ab initio MD, or large-scale models alone. More broadly, the approach can be used to evaluate the behavior of the large-scale pretrained MLIP on timescales it cannot directly access.

\begin{figure*}[t]
  \centering
  \includegraphics[width=\textwidth]{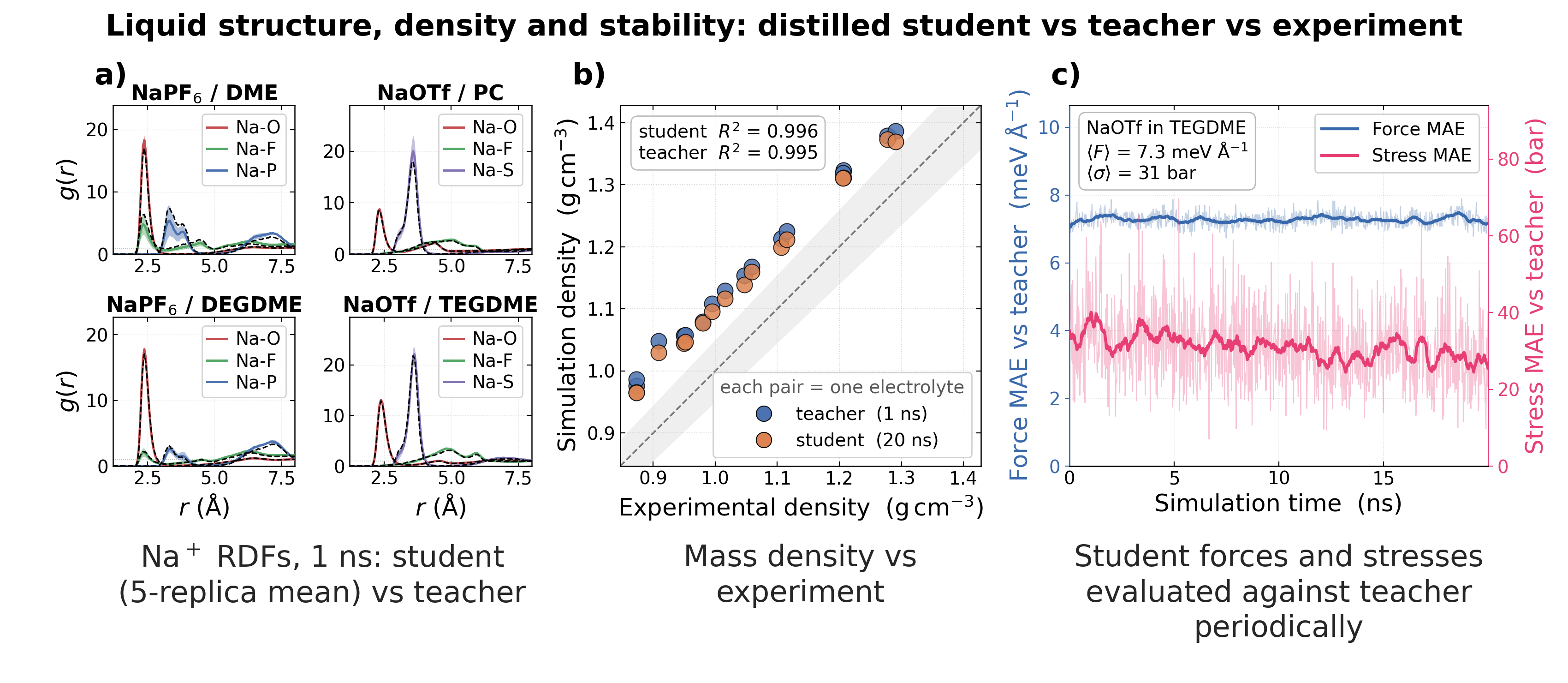}
    \caption{\textbf{Distilled PaiNN potentials reproduce the static observables of the UMA-OMol teacher at short timescales.} \textbf{(a)}~Radial distribution functions of four electrolyte systems across multiple correlation pairs, obtained from 1~ns trajectories at 293~K, with both the teacher model (UMA-OMol) and the student (PaiNN) displayed. Each PaiNN line is the mean over five independent 1~ns replicas, where the shading indicates the $\pm 1$~s.d.\ replica spread. The teacher curve falls within the shaded band for almost every pair and system, indicating that the teacher lies within the student's per-replica noise and that the student accurately reproduces teacher behavior. \textbf{(b)}~Parity plot of simulated versus experimentally measured mass density, where each dot denotes a different electrolyte system. Both UMA-OMol and PaiNN show a systematic upward bias against experiment, but UMA-OMol and PaiNN coincide for almost every system, so distillation adds little error of its own. The individual densities for each electrolyte are listed in Sec.~\ref{si:density}. \textbf{(c)}~For a 20~ns \emph{NPT} 1~M NaOTf in TEGDME simulation, we evaluate force mean absolute error (MAE) and stress MAE relative to UMA-OMol at 20~ps intervals by using UMA-OMol to compute single-point estimates. Light traces are per-frame errors, while bold curves show the $\sim$500~ps running average. Both errors stay low, with no upward drift over 20~ns: a linear fit changes the force MAE by $<$0.2\%, and the stress MAE decreases slightly over time. Similar plots for the production runs are shown in Sec.~\ref{si:stability}.}
\label{fig:fidelity}
\end{figure*}

\section*{Results}

\begin{figure*}[t]
  \centering
  \includegraphics[width=0.90\textwidth]{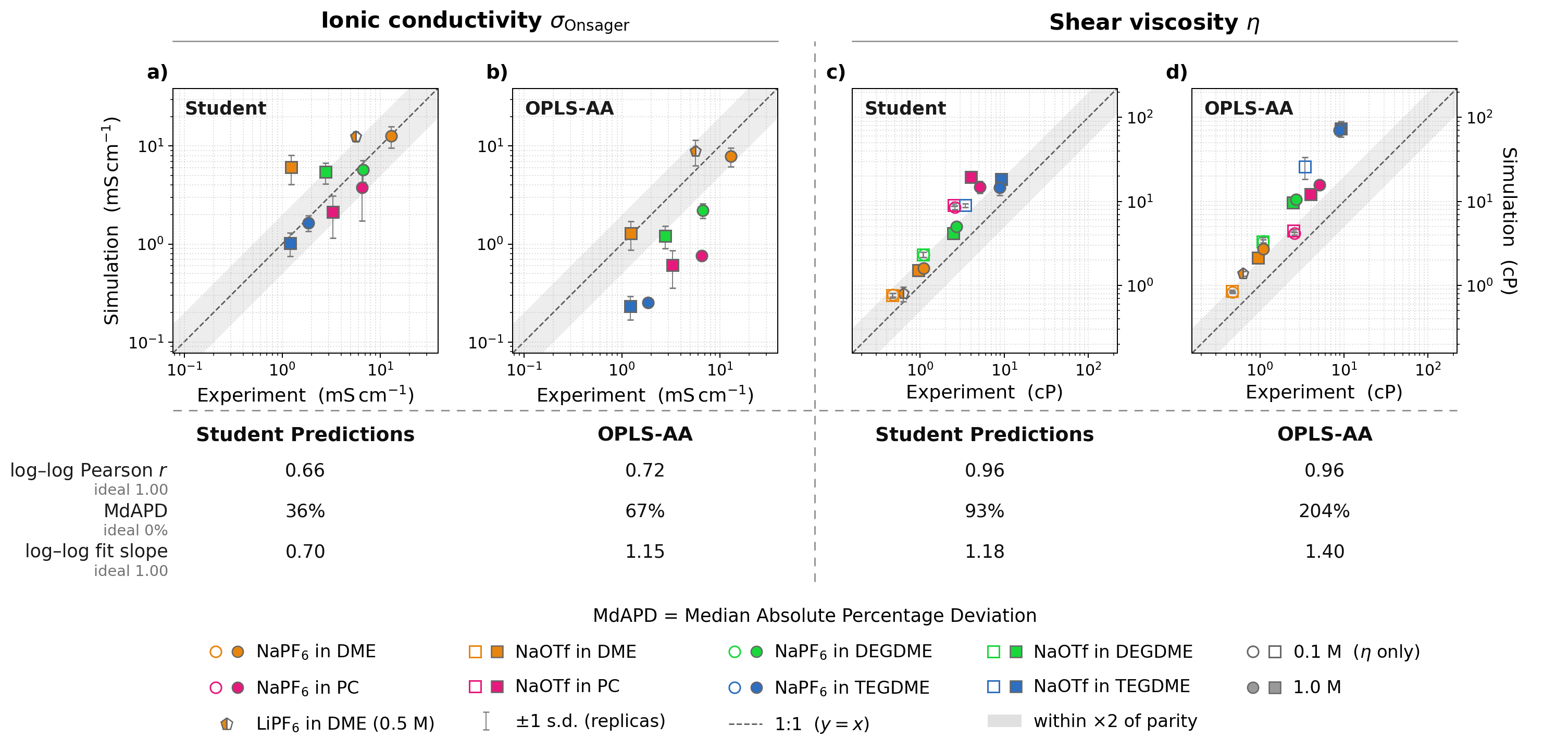}
  \caption{\textbf{Ionic conductivity and shear viscosity vs.\ experiment at 298~K, for the distilled student models and the OPLS-AA force field.} Each point is the mean over four independent 20~ns replicas, and error bars show $\pm 1$~s.d.\ across the replicas. \textbf{(a,b)}~Ionic conductivity $\sigma$ for the PaiNN student and OPLS-AA. Conductivity is only computed for the concentrated ($\geq$0.5~M) systems, since at 0.1~M the box holds too few charge carriers to converge the collective estimator. The student roughly halves the median absolute percentage deviation of OPLS-AA (MdAPD; 36\% vs.\ 67\%), with a comparable log--log correlation. \textbf{(c,d)}~Viscosity $\eta$ against experiment for the two force fields. Again the student tracks experiment much more closely than OPLS-AA, with an MdAPD less than half that of OPLS-AA and a log--log fit slope closer to one.}
  \label{fig:conductivity}
\end{figure*}

\begin{figure*}[t]
  \centering
  \includegraphics[width=0.75\textwidth]{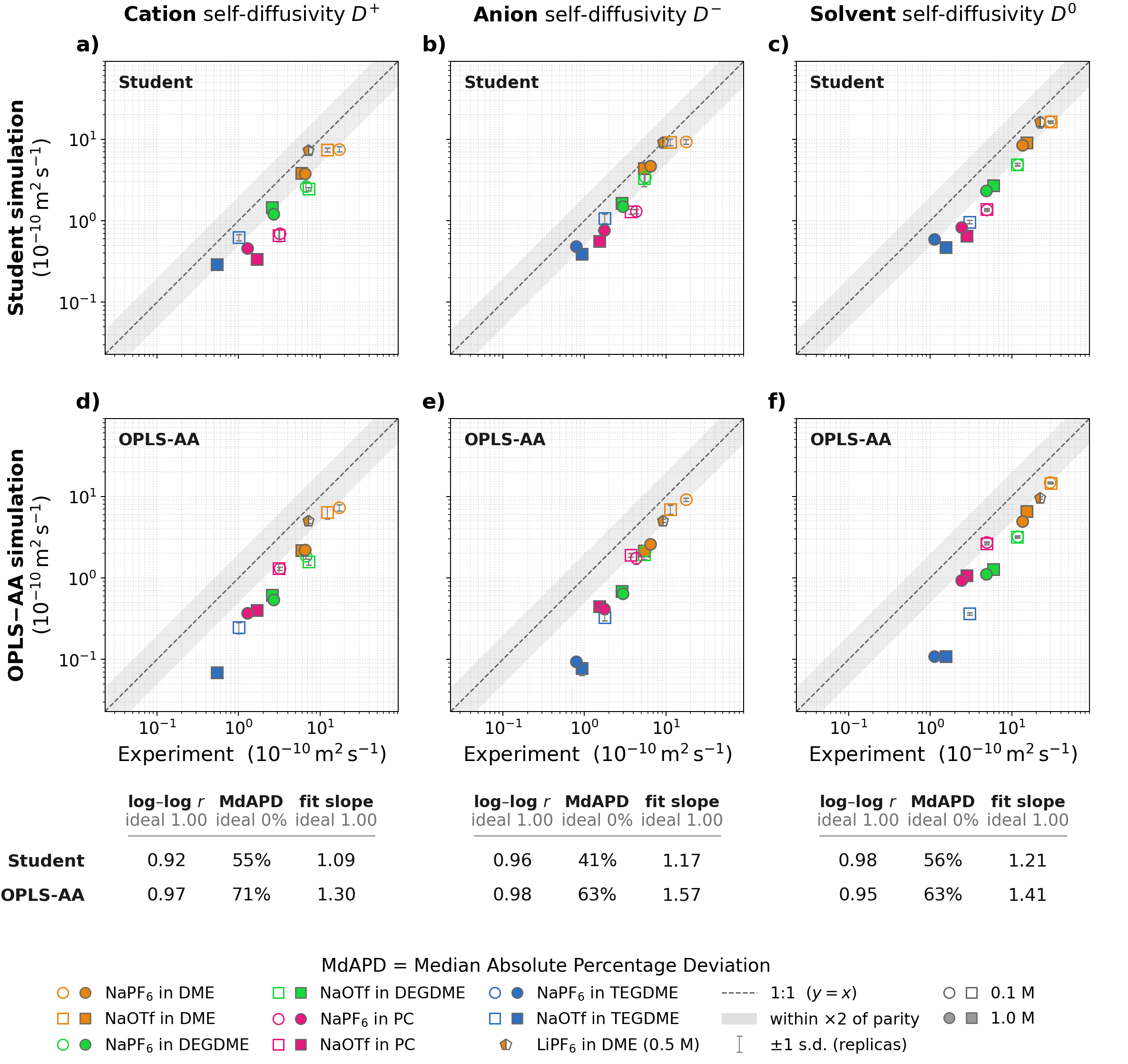}
  \caption{\textbf{Self-diffusivities vs.\ experiment at 298~K, for the distilled students and the OPLS-AA force field.} Self-diffusivities of the cation $D^+$ \textbf{(a,d)}, anion $D^-$ \textbf{(b,e)}, and solvent $D^0$ \textbf{(c,f)}, for the PaiNN student (top row) and OPLS-AA (bottom row), where each point is the mean over four independent 20~ns replicas (error bars, $\pm 1$~s.d.\ across the replicas), with the Yeh--Hummer finite-size correction applied \cite{yehhummer2004}. The student outperforms OPLS-AA for every species, both in MdAPD and in log--log fit slope.}
  \label{fig:diffusivity}
\end{figure*}

The distilled PaiNN potentials enable simulations long enough to converge transport coefficients, tracking experiment across sixteen electrolyte systems. We first evaluate the student's faithfulness at the 1~ns timescale accessible to the teacher, by comparing both models' radial distribution functions and mass densities. We additionally check the student's per-frame force and stress errors periodically across 20~ns of student simulation. The students' ${\sim}40\times$ speed advantage enables 20~ns simulations, from which we compute ionic conductivities, shear viscosities, and self-diffusivities. Student-predicted transport properties, benchmarked against experiment, outperform the OPLS-AA classical force field \cite{jorgensen1996}. The sixteen systems are sodium electrolytes combining NaPF$_6$ or NaOTf with monoglyme (DME), diglyme (DEGDME), tetraglyme (TEGDME), or propylene carbonate (PC) at 0.1 and 1~M (NaPF$_6$/TEGDME at 1~M only), plus a lithium baseline, LiPF$_6$ in DME at 0.5~M. Full compositions are listed in Table~\ref{tab:systems}; details of the box setup, MD ensemble settings, and student training are given in the Methods and Secs.~\ref{si:systems}--\ref{si:training} of the Supplementary Information (SI).

\subsection*{Reproducing the behavior of the teacher model}

At the 1~ns timescales accessible to the teacher, the student reproduces both its radial distribution functions (RDFs) and its mass density. The RDF curves of the student and teacher overlap across all pair types, and the agreement holds across all 1~M systems (full set in Sec.~\ref{si:rdf}), indicating that the student accurately captures the teacher's first-shell coordination as well as its longer-range features (Fig.~\ref{fig:fidelity}a). The distilled model also matches UMA-OMol's mass densities, agreeing with UMA-OMol to within 1.1\% on average across all sixteen electrolytes (Fig.~\ref{fig:fidelity}b). Both models overestimate the experimental densities by 6--15\% (about 10\% on average); however, this is a known tendency of UMA-OMol \cite{kumar2026}, and distillation adds little error to the existing bias.

The student additionally maintains low force and stress errors relative to the teacher throughout long simulations, as a result of augmenting the training data with rattled structures. For a representative system (1~M NaOTf in TEGDME), we perform single-point UMA-OMol force and stress evaluations at 20~ps intervals along a student 20~ns NPT trajectory. This serves as a low-cost spot check on the student's accuracy that avoids running the teacher itself for the full 20~ns. The force mean absolute error (MAE) against UMA-OMol is around $7$~meV/\AA{} and the stress MAE near $30$~bar. Both errors are low with no upward drift (Fig.~\ref{fig:fidelity}c). The force errors stay similarly low along the NVT production trajectories used for the transport coefficients, for every system (Figs.~\ref{fig:stability-01M} and~\ref{fig:stability-1M}). The student trajectory's stability is a direct result of adding rattled structures to the training set. We isolate the effect of including noised data during training in Sec.~\ref{si:noise}. Trained without noised data, the student quickly wanders out of the narrow configurational region its training covers. The student's force error grows as the simulation progresses, until the system destabilizes within a nanosecond. Exposing the student to off-equilibrium configurations teaches it a restoring force that corrects small deviations and prevents instability from compounding.

\subsection*{Computing transport properties with the student model}

The distilled models' ${\sim}40\times$ speedup over the teacher enables converged transport properties that track experimental trends and consistently outperform OPLS-AA, the classical force-field baseline. Using the distilled potentials, we run 20~ns production simulations and extract ionic conductivities, shear viscosities, and self-diffusivities, which are then compared to experiment (Figs.~\ref{fig:conductivity} and~\ref{fig:diffusivity}). Every reported coefficient is the mean over four 20~ns NVT replicas launched from a common configuration with independently seeded velocities. Along each trajectory, we confirm that the energy, temperature, and pressure are stable, and that the force error against the teacher remains low for the full 20~ns (Sec.~\ref{si:stability}). For each property, we evaluate model performance against experiment using three metrics. A log--log Pearson correlation coefficient ($r$) near unity indicates that predictions tightly track experiment; a low median absolute percentage deviation (MdAPD) indicates small typical errors; and a log--log linear regression slope near one indicates that the model reproduces the experimental spread across systems.

Computed student conductivities track experiment across over an order of magnitude, and possess a lower MdAPD than OPLS-AA. Results are reported for the nine concentrated ($\geq$0.5~M) systems; at 0.1~M the box contains too few charge carriers for the collective estimator to converge. Seven of the nine systems sit within a factor of two, and the model reaches an MdAPD of 36\% (Fig.~\ref{fig:conductivity}a). OPLS-AA is notably less accurate, as it underpredicts seven of the nine conductivities, places only three within the factor-of-two band, and has an MdAPD of 67\%. Its log--log correlation ($r = 0.72$) is comparable to the students' ($0.66$) (Fig.~\ref{fig:conductivity}b). The student does have a large error for 1~M NaOTf in DME, overpredicting the conductivity by a factor of five, unlike OPLS-AA. This system has the lowest ionicity of the concentrated systems, and in Sec.~\ref{si:conductivity} we use ion-pair scans to show that this error likely arises from the teacher's finite 6~\AA{} graph cutoff, which causes ions to pair less than they should.

Student viscosities likewise track the experimental ranking closely; their consistent upward bias is attributable to the teacher's systematic density overshoot. Because the students inherit the teacher's overestimate of the liquid densities ($6$--$13\%$ for the students; Fig.~\ref{fig:fidelity}b), the simulated liquids are all over-compressed compared to experiment. Most of the reduction in space comes out of the free volume in the box (the volume not occupied by the molecules) \cite{doolittle1951,cohenturnbull1959}. In Sec.~\ref{si:visc-density}, using high-pressure viscometry data of comparable liquids from the literature, we estimate that a density overshoot of this size should inflate our predicted viscosity by roughly three- to fourfold. Consistent with this, every student viscosity lies above experiment, by $2.1\times$ on geometric average (Fig.~\ref{fig:conductivity}c). Student predictions are tightly correlated with experiment ($r = 0.96$) and mostly follow a linear trend, with a log--log slope of $1.18$. OPLS-AA likewise overpredicts the viscosity, but much more strongly, with an MdAPD of 204\% against the students' 93\% (Fig.~\ref{fig:conductivity}d). OPLS-AA's overprediction is also much less uniform, growing as the viscosity itself grows (log--log fit slope $1.40$), from under $2\times$ for the least viscous systems to about $8\times$ the experimental value for the two 1~M TEGDME systems.

Self-diffusivities follow the same pattern, with the distilled models reproducing experimental trends more faithfully than the classical baseline (Fig.~\ref{fig:diffusivity}). The student models achieve $r = 0.92$--$0.98$ and MdAPDs of 55\%, 41\%, and 56\% for $D^+$, $D^-$, and $D^0$. The students underpredict the diffusivities by roughly $2\times$, which mirrors the overshoot we saw for viscosity: by the approximate Stokes--Einstein relation $D \propto k_{\mathrm{B}}T/\eta$, a $2\times$ increase in viscosities should result in a decrease in diffusivities by roughly the same factor, which is what we observe. Both the viscosity overprediction and diffusivity underprediction are likely caused by the density bias inherited from the teacher, not by shortcomings introduced by distillation. OPLS-AA performs notably worse, underpredicting the diffusivities by roughly a factor of three with MdAPDs of 63--71\% (Fig.~\ref{fig:diffusivity}d--f). Both models achieve comparable log--log correlations, but the distilled models follow a much more linear relationship against experiment, with log--log slopes of $1.09$--$1.21$ compared with $1.30$--$1.57$ for OPLS-AA.

Beyond the absolute values, the students recover expected chemical trends. Self-diffusivities decrease with both salt concentration and glyme chain length (DME $>$ DEGDME $>$ TEGDME). The solvent moves fastest in each electrolyte, and the predicted ordering of the cation and anion diffusivities agrees with experiment, except in cases where the two are nearly equal.

The transport properties reported here required thoroughly sampled MD simulations; trajectories limited to 1--5~ns were insufficient to converge the coefficients (Sec.~\ref{si:convergence}). Critically, using the teacher to simulate these trajectories would be impractical; at UMA-OMol throughput ($\sim$0.23~ns/day per system, Sec.~\ref{si:speed}), each 20~ns replica would take roughly three months to finish. In contrast, the PaiNN student models complete 20~ns in about two days.

\section*{Discussion}

The student's combination of speed and accuracy makes large-scale pretrained MLIPs practical for transport screening. A new electrolyte requires only a short teacher trajectory, roughly ten GPU-hours of student training, and about two days of production MD to obtain transport coefficients. Although each student is specific to a single salt--solvent pair, the cost of training is small compared to the time saved by using the student for inference. Because OMol25 covers a broad range of molecular chemistry, the teacher can be applied to new electrolytes beyond the systems studied here, without additional ab initio data. 

The student's fidelity to the teacher bounds the accuracy of its predictions, but it also provides a new way to evaluate the long-timescale behavior of large-scale pretrained MLIPs. UMA-OMol overestimates the densities of our electrolytes by about 10\% on average, and distillation reveals that this translates into a systematic $\sim$2$\times$ viscosity overshoot and a corresponding $\sim$2$\times$ diffusivity undershoot. Without access to the long trajectories produced by the student, we would not know whether this error was due to statistical fluctuations or an actual shortcoming of the teacher model. The student's inheritance of teacher shortcomings diminishes accuracy against experiment; however, this is a transient limitation. Large-scale pretrained models are quickly improving, with training datasets growing in diversity and new architectures incorporating explicit long-range electrostatics \cite{mace_polar} that directly remove the cutoff artifact identified in Sec.~\ref{si:conductivity}. Our pipeline consumes only the energies, forces, and stresses the teacher assigns, and is agnostic to the choice of architecture or the data the teacher was trained on. Every improvement to the large-scale model therefore propagates to the distilled transport predictions, without any change to the method.

Large-scale pretrained models have been shown to reproduce the structure of liquid electrolytes; we show distillation extends them to electrolyte dynamics at scale. Converged transport coefficients at the accuracy of large-scale models have so far required a months-long computation per system. Knowledge distillation reduces this cost enough that such calculations could be incorporated into the screening loops needed for next-generation electrolyte design.

\section*{Methods}

\subsection*{Electrolyte systems and simulation cells}

We study sixteen systems in this work: NaPF$_6$ and NaOTf in monoglyme (DME), diglyme (DEGDME), tetraglyme (TEGDME), and propylene carbonate (PC) at 0.1 and 1~M (NaPF$_6$ in TEGDME at 1~M only), plus LiPF$_6$ in DME at 0.5~M; per-system compositions and box sizes are listed in Table~\ref{tab:systems}. The final boxes for production molecular dynamics (MD) runs contain 2700--3700 atoms. For MD simulations, we prepare initial configurations using an earlier structural study's protocol \cite{kumar2026}. Specifically, we use PACKMOL to place molecules into cubic cells that are 30~\AA{} on each side~\cite{packmol}, perform geometric relaxation, then equilibrate each box in the isothermal--isobaric (NPT) ensemble with the OPLS-AA force field \cite{jorgensen1996} for 1~ns to relax the initial packing. Because the boxes are packed at a non-equilibrium volume, the molarity drifts significantly from its target when the barostat relaxes. To mitigate this, we use each system's simulated equilibrium density to recompute the correct solute-to-solvent ratio, and repack the box. The corrected boxes equilibrate to within 1.4\% of the target molarity on average. We provide the derivation of our correction, as well as per-system shifts, in Sec.~\ref{si:systems}.

\subsection*{Teacher reference data}
Reference data for training the distilled models is generated with UMA-OMol (checkpoint \texttt{uma-s-1p1}) using its OMol task head \cite{uma, omol25}. For each system and concentration we run a 1~ns trajectory in the NPT ensemble at 293~K and 1~bar using the Atomic Simulation Environment (ASE) \cite{ase}, with a Nos\'e--Hoover thermostat (coupling time 100~fs) \cite{nose1984,hoover1985}, a Parrinello--Rahman-type barostat~\cite{parrinello1981}, and a 1~fs timestep. We then sample the first 100~ps of each trajectory to make the training configurations; the last 70\% of the trajectory is used as the reference for the radial distribution function comparison between student and teacher.

Unperturbed snapshots are sampled every 50~fs from the teacher trajectory. A second set of snapshots, sampled every 50~fs but offset by 20~fs from the first grid, is collected and then perturbed by adding Gaussian noise with a standard deviation of 0.1~\AA{} to every atom. The simulation cell is not perturbed. Previous work has found that augmenting training with Gaussian-noised structures improves the robustness of MLIPs \cite{yuan2025teachers}. The noised points allow the model to see a wider region of configuration space, and help stabilize the distilled potential during simulations. We isolate the effect of the noised data in Sec.~\ref{si:noise}. Every configuration is labeled with UMA-OMol energies, forces, and the full stress tensor. We use a 90:10 train--validation split, giving 3600 training and 400 validation configurations per concentration. When two concentrations exist for a salt--solvent pair, we combine them into one dataset so that a single potential is trained across both loadings.

\subsection*{Student potential and training}

We use the PaiNN equivariant message passing graph neural network~\cite{painn} for our student model, as implemented in the FeNNol repository in JAX \cite{fennol}. The model has three message passing layers, an embedding dimension of 128, and a radial neighbor cutoff of 5.1~\AA{}. We also add a trainable Ziegler--Biersack--Littmark (ZBL) term \cite{zbl} with an inner cutoff of 1.1~\AA{} for explicit short-range nuclear repulsion. To train the model, we minimize the loss
\begin{equation}
\label{eq:loss}
\mathcal{L} = \frac{\lvert E - \hat{E}\rvert}{N} + \frac{1}{N}\sum_{i=1}^{N}\lVert \mathbf{F}_i - \hat{\mathbf{F}}_i \rVert + \frac{1}{9}\sum_{\alpha\beta}\lvert \sigma_{\alpha\beta} - \hat{\sigma}_{\alpha\beta}\rvert,
\end{equation}
 where $N$ is the number of atoms and hats denote UMA-OMol labels. We use AdamW \cite{adamw} as the optimizer and train for 275 epochs with a one-cycle cosine learning-rate schedule (peak learning rate $8 \times 10^{-4}$). Training a single model takes 9--22 hours on one NVIDIA GH200 GPU, depending on the number of pooled concentrations (Table~\ref{tab:datasets}). The full architecture and optimization settings are given in Table~\ref{tab:arch}.

\subsection*{Structural validation against the teacher}
Each distilled potential is validated against the teacher using radial distribution functions and mass densities, which are short timescale observables. Between species $A$ and $B$, the radial distribution function is defined as 
\begin{equation}
\label{eq:rdf}
g_{AB}(r) = \frac{1}{N_A\,\rho_B}\left\langle \sum_{i\in A}\sum_{j\in B} \frac{\delta(r - r_{ij})}{4\pi r^2} \right\rangle,
\end{equation}
where $r_{ij}$ is the minimum-image distance between atoms $i$ and $j$, $N_A$ is the number of $A$ atoms, and $\rho_B = N_B/V$ is the bulk number density of $B$. For each system, we run five independent 1~ns student NPT simulations at 293~K and compare them with a matched 1~ns UMA-OMol run. Each RDF is computed identically, using 3000 configurations sampled from the last 70\% of the trajectory, with a 0.05~\AA{} bin width out to 12~\AA{} (Sec.~\ref{si:rdf}). The 0.1~M cells only contain two ion pairs, which is not enough to converge the RDFs in 1~ns of simulation. We calculate the students' mass densities from a 20~ns NPT trajectory at 298~K, averaged after discarding the first 0.5~ns, and obtain the corresponding UMA-OMol densities from our own 1~ns NPT simulations at 298~K with the teacher checkpoint. 

\subsection*{Production simulations}
We run production simulations with the student model using the JAX-based MD engine implemented in FeNNol \cite{fennol,jax}. Each molarity-corrected box (Sec.~\ref{si:systems}) is first equilibrated for 1~ns in an NPT ensemble at 298~K and 1~bar, using a Langevin thermostat (friction 1~ps$^{-1}$) and an isotropic Langevin-piston barostat (piston relaxation time 200~fs). We compute the mean density over the last 500~ps of this trajectory and select the frame from the last 300~ps whose instantaneous density lies closest to the mean. This frame is then used as the starting configuration of the production simulation (Sec.~\ref{si:nvt-seeds}).

The four 20~ns NVT production replicas run at 298~K, each starting with different initial Maxwell--Boltzmann velocities and thermostat noise. We use the global Bussi (CSVR) velocity-rescaling thermostat \cite{bussi2007} with a 10~ps relaxation time. Holding the volume fixed removes barostat fluctuations from the stress autocorrelations, and the slow global thermostat leaves sub-picosecond correlations unbiased \cite{basconi2013,maginn2019}. We use a 1~fs timestep and save frames every 1~ps and the stress tensor used to calculate the shear viscosity every 2~fs. All MD settings are reported in Table~\ref{tab:mdsettings}.

To verify that the distilled potentials remain consistent with UMA-OMol over these long trajectories, we perform single-point UMA-OMol evaluations on 500 frames per student trajectory, spaced 40~ps apart across the full 20~ns, and monitor the mean and maximum force errors throughout the simulation, together with other observables (potential energy per atom, temperature, and pressure). We observe low and flat force errors for every system, and the other diagnostic observables are very stable throughout each trajectory (Sec.~\ref{si:stability}).

\subsection*{Computational throughput}
We test the throughput of both student and teacher models on a single NVIDIA GH200. The student reaches a median of 9.5~ns/day across the benchmarked boxes versus 0.23~ns/day for UMA-OMol, a 41$\times$ speedup. Per-system throughputs are reported in Sec.~\ref{si:speed}.

\subsection*{Classical force-field baseline}

We use GROMACS 2025.3 \cite{gromacs} to run simulations with the OPLS-AA classical force field \cite{jorgensen1996}. Solvent parameters (DME, DEGDME, TEGDME, PC) are OPLS-AA with 1.14*CM1A charges from LigParGen \cite{ligpargen}. Both anions use the CL\&P parameter set, which is built on the OPLS-AA functional form: PF$_6^-$ from Canongia Lopes et al.~\cite{canongialopes2004pf6} with the bonded terms of Kaminski and Jorgensen \cite{kaminski1999}, and OTf$^-$ from Canongia Lopes and P\'adua \cite{canongialopes2004otf} with the triflate oxygen charge of Gouveia et al.~\cite{gouveia2017}. Na$^+$ and Li$^+$ use the \AA qvist parameters \cite{aqvist1990} as distributed with GROMACS. Lennard-Jones cross terms follow geometric combination rules and 1--4 interactions are scaled by 0.5, as in OPLS-AA. All ions carry their full formal charge. The parameter sources are summarized in Table~\ref{tab:oplssettings}, and the full per-species parameters are provided with the released code. We correct the molarity of the initial boxes with the same procedure as for the student model, but with OPLS-AA's own equilibrium density. Steepest-descent energy minimization is used to relax the packed boxes, and the boxes are then equilibrated for 1~ns of NPT at 298~K and 1~bar (using a velocity-rescaling (CSVR) thermostat with a 0.1~ps relaxation time and an isotropic stochastic cell-rescaling (C-rescale) barostat \cite{bernetti2020} with a 2~ps relaxation time). Finally, we run four independent 20~ns NVT production replicas at the equilibrated volume; the full settings are tabulated in Table~\ref{tab:oplssettings}. During the production run, we use a 1~fs timestep with no bond constraints (fully flexible molecules), and each replica is initialized with Maxwell--Boltzmann velocities from an independent seed. Particle-mesh Ewald \cite{essmann1995} is used for the electrostatics with a 12~\AA{} real-space cutoff, and the Lennard-Jones interactions are cut at 12~\AA{} with an analytic long-range dispersion correction. We save frames every 1~ps and the pressure tensor every 2~fs.

\subsection*{Experimental measurements}

All electrolytes were prepared and handled inside an argon-filled glovebox (O\textsubscript{2} below 0.2~ppm, H\textsubscript{2}O below 0.1~ppm). The experimental densities, self-diffusivities, ionic conductivities, and shear viscosities were measured in-house; the values are tabulated in Sec.~\ref{si:transportdata}. The densities were also used in an earlier structural study \cite{kumar2026}. Self-diffusivities were measured by pulsed-field-gradient nuclear magnetic resonance (PFG-NMR) on 11.7~T Bruker Avance NEO spectrometers, following \textsuperscript{1}H for the solvent, \textsuperscript{7}Li or \textsuperscript{23}Na for the cation, and \textsuperscript{19}F for the anion; the most viscous \textsuperscript{23}Na samples were measured on a 14.1~T Avance III HD system with a Diff50 probe. Ionic conductivities were measured in the glovebox with a Mettler Toledo SevenDirect SD30 meter and an InLab 710 four-electrode platinum probe, with three independent readings per electrolyte. Density and viscosity were obtained from a single sample filling at 298~K with an Anton Paar DMA 4500 M oscillating U-tube density meter coupled to a Lovis ME rolling-ball viscometer, both calibrated with certified reference water. Full protocols for all four measurements are given in Sec.~\ref{si:experiment}.

\subsection*{Transport-property analysis}

All transport coefficients, simulated and experimental, are reported at 298~K. Simulated coefficients are computed independently for each production replica. To alleviate the noise introduced by early-stage equilibration, we discard the first 2~ns of each trajectory and use the remainder for the computation. Results are reported as the mean and standard deviation over the four replicas.

\subsubsection*{Self-diffusivity}
We calculate self-diffusivity by fitting the slope of the mean-squared displacement (MSD) over time lag $\tau$ based on the Einstein relation \cite{einstein1905,hansen2013,jones2012},
\begin{equation}
D = \frac{1}{6}\lim_{\tau\to\infty}\frac{d}{d\tau}\big\langle|\mathbf{r}(t_0+\tau)-\mathbf{r}(t_0)|^2\big\rangle ,
\end{equation}
where $t_0$ is the time origin, and $\langle\cdots\rangle$ denotes an ensemble average evaluated over all available time origins as well as over all particles of the species of interest (specifically, cation, anion, and solvent).

We first compute the unwrapped coordinates of each trajectory using the minimum-image convention, then subtract the center-of-mass drift of the entire system. We use the molecular centers of mass for the per-species displacements. MSDs across different time lags are calculated using a fast Fourier transform (FFT) and fitted following the best practices of Maginn \emph{et al.}\ \cite{maginn2019}. Instead of using a fixed lag window, for each species we evaluate the logarithmic slope $\nu(\tau) = \mathrm{d}\ln \mathrm{MSD} / \mathrm{d}\ln\tau$ on logarithmically spaced lags between 5~ps and 36\% of the maximum lag. We then fit the MSD over the longest contiguous run satisfying $|\nu - 1| \le 0.15$, requiring at least twelve points. If no such window exists, we perform the fit over the twelve-point stretch with the smallest $|\nu - 1|$ value.

After calculating the periodic-box diffusivity $D_{\mathrm{PBC}}$ from the fitted slope, we apply the Yeh--Hummer correction to remove finite box size effects \cite{yehhummer2004}, $D = D_{\mathrm{PBC}} + \Delta D$, with
\begin{equation}
\Delta D = \frac{k_{\mathrm{B}} T\, \xi}{6\pi\,\eta\, L},
\end{equation}
where $\xi = 2.837297$ for a cubic cell, $L$ is the box edge length, and $k_{\mathrm{B}} T$ is the thermal energy. For $\eta$ we use the Green--Kubo viscosity of the same system, averaged over its replicas, so the correction is fully self-consistent and involves no experimental input. Both the uncorrected diffusivities and the corrections are listed in Tables~\ref{tab:transport-diffusivity-painn} and~\ref{tab:transport-diffusivity-opls}.

\subsubsection*{Ionic conductivity}
The ionic conductivity is calculated from the Onsager transport coefficients $L_{ij}$ as \cite{fong2020,fong2020macro},
\begin{equation}
\label{onsager_conductivity}
\sigma = e^2\sum_{ij}z_i z_j L_{ij},
\end{equation}
where $e$ is the elementary charge and $z_i$ is the signed integer charge of species $i$. Each ion is treated as a point charge at its center of mass, with $z=+1$ for the Na$^+$ and Li$^+$ cations, $z=-1$ for the PF$_6^-$ and OTf$^-$ anions, and $z=0$ for the solvent molecules. We first define the cumulative displacement of all species-$i$ particles over a lag $\tau$ as
\begin{equation}
\label{onsager_dR}
\Delta\mathbf{R}_i(t_0,\tau) = \sum_{k=1}^{N_i}\left[\mathbf{r}_{ik}(t_0+\tau) - \mathbf{r}_{ik}(t_0)\right],
\end{equation}
where $\mathbf{r}_{ik}$ is the center-of-mass position of the $k$-th molecule of species $i$, measured relative to the system's center of mass, and $N_i$ is the total number of species-$i$ particles in the simulation box. We next define the cross-correlation of species $i$ and $j$ as 
\begin{equation}
\label{onsager_cij}
C_{ij}(\tau) = \left\langle \Delta\mathbf{R}_i(t_0,\tau) \cdot \Delta\mathbf{R}_j(t_0,\tau) \right\rangle .
\end{equation}
The angle brackets $\langle\cdots\rangle$ denote an ensemble average, which is evaluated as an average over all available time origins $t_0$. The transport coefficient is then given by the long-time slope of the cross-correlation $C_{ij}(\tau)$ as
\begin{equation}
\label{onsager_lij}
L_{ij} = \frac{\beta}{6V}\,\lim_{\tau\to\infty}\frac{\partial}{\partial \tau}C_{ij}(\tau),
\end{equation}
where $\beta = 1/(k_\mathrm{B}T)$ and $V$ is the volume of the simulation box.  The slope is extracted by a least-squares fit over lag times of 0.5--2~ns. We use a shorter window than for the self-diffusivities, because the displacement for conductivity is collective and does not include particle averaging, making noise a significant issue at longer lags. Unlike the Nernst--Einstein estimate, this correlated estimator retains the ion--ion cross-correlations \cite{fong2020}; the derivation and the comparison between the two estimators are given in Sec.~\ref{si:conductivity}. Conductivities are reported for the nine concentrated ($\geq$0.5~M) systems only: at 0.1~M the cell contains only two ion pairs, which is not enough to accurately measure distinct-pair (cross) correlations in $L_{ij}$.

\subsubsection*{Shear viscosity}
The shear viscosity $\eta$ is computed from the equilibrium Green--Kubo stress-autocorrelation function \cite{green1954,kubo1957,hansen2013,j2007statistical, jones2012},
\begin{equation}
  \eta \;=\; \frac{V}{k_{\mathrm B}T}
  \int_{0}^{\infty}
  \bigl\langle\, \sigma_{\alpha\beta}(0)\,\sigma_{\alpha\beta}(t) \,\bigr\rangle
  \;\mathrm{d}t,
  \label{eq:eta_gk}
\end{equation}
where $\sigma_{\alpha\beta}$ is an off-diagonal component, such as $\sigma_{xy}$, of the instantaneous stress tensor, logged along the trajectory. To reduce statistical noise, we average the autocorrelation in Eq.~\eqref{eq:eta_gk} over five shear components of the traceless symmetric stress tensor, each of which integrates to the same value $\eta$ \cite{daivis1994,maginn2019}:
\begin{equation}
  \sigma_{xy},\quad \sigma_{xz},\quad \sigma_{yz},\quad
  \tfrac{1}{2}(\sigma_{xx}-\sigma_{yy}),\quad
  \tfrac{1}{2}(\sigma_{yy}-\sigma_{zz}).
  \label{eq:eta_comps}
\end{equation}
The autocorrelation is evaluated by FFT and integrated by the trapezoidal rule to give the running estimate $\eta(t)$, accumulated over lags up to 1~ns. We use the time-decomposition method described by Zhang \emph{et al.}\ \cite{zhang2015viscosity,maginn2019} to obtain the reported value. The integral is fit with a double-exponential form over a window that ends at the lag $t_{\mathrm{cut}}$, defined as the first point at which the standard deviation across five equal time blocks of the trajectory exceeds 40\% of the running value. Each fit point is weighted inversely by the standard deviation. We extrapolate our fit to infinite time to get $\eta_\infty$, and this value is reported. This procedure prevents the noisy tail of the integral from biasing the estimate. Convergence checks and example fits are given in Sec.~\ref{si:viscosity-details}.

\subsection*{Use of large language models}
Large language models (Claude, Anthropic) were used to assist with writing code. All code was thoroughly reviewed by the authors, who take full responsibility for the content.

\section*{Data availability}
The trained student models, the initial and molarity-corrected simulation boxes, and the per-replica transport analysis curves are available at \url{https://github.com/ASK-Berkeley/distillation-electrolytes}. The molecular dynamics trajectories and distillation datasets can be regenerated with the released code, models, and starting configurations.

\section*{Code availability}
All code used in this work is available at \url{https://github.com/ASK-Berkeley/distillation-electrolytes}. The student models are trained and simulated with a fork of FeNNol~\cite{fennol}, included in the repository as a submodule.

\section*{Acknowledgements}
We would like to thank Kranthi K. Mandadapu for the guidance, help, and discussion on the theory behind transport property calculation and error quantification in this paper. We would also like to thank Bryan D. McCloskey for allowing us to use his experimental equipment to make measurements for the work.

We would also like to thank Sakib Matin, Sanjeev Raja, and Eric Qu for insightful discussions that benefited the work.

This work used the DeltaAI GPU resource at the National Center for Supercomputing Applications through allocation BIO250235 from the Advanced Cyberinfrastructure Coordination Ecosystem: Services \& Support (ACCESS) program~\cite{access2023}, which is supported by U.S. National Science Foundation grants \#2138259, \#2138286, \#2138307, \#2137603, and \#2138296. This research used the DeltaAI advanced computing and data resource, which is supported by the National Science Foundation (award OAC 2320345) and the State of Illinois. DeltaAI is a joint effort of the University of Illinois Urbana-Champaign and its National Center for Supercomputing Applications.

\section*{Funding}
This work was partially funded by the Office of Science, Office of Basic Energy Sciences in the Chemical Sciences, Geosciences, and Biosciences subprogram at Lawrence Berkeley National Laboratory under contract \#DE-AC02-05CH11231. We also acknowledge funding from the Energy Storage Research Alliance (ESRA) (DE-AC02-06CH11357), an Energy Innovation Hub funded by the U.S. Department of Energy, Office of Science, Basic Energy Sciences.  

\section*{Author contributions}
I.A. and Y.J. formulated the distillation procedure and ran all model training, simulations, and transport calculations. N.K. helped supply the initial boxes for the work, as well as code for static observables. C.Z. helped train and simulate initial versions of distilled models. J.L. prepared the electrolytes and provided the salts and solvents. C.S.M. and J.D.B. conducted the density, viscosity, and self-diffusivity measurements. A.M.M. and J.R.T. performed ionic conductivity measurements. E.J.C., J.D.B., and N.P.D. supervised the experimental work. N.K., I.A., Y.J., S.B., and A.S.K. conceptualized and designed the study. A.S.K. and S.B. supervised the work. I.A., Y.J., N.K., S.B., and A.S.K. all helped write up the final manuscript.

\section*{Competing interests}
The authors declare no competing interests.

\newpage
\bibliography{citation}
\IfFileExists{appendix.pdf}{%
  \clearpage\onecolumngrid\pagestyle{empty}%
  \ifdefined\pdflastximagepages
    \pdfximage{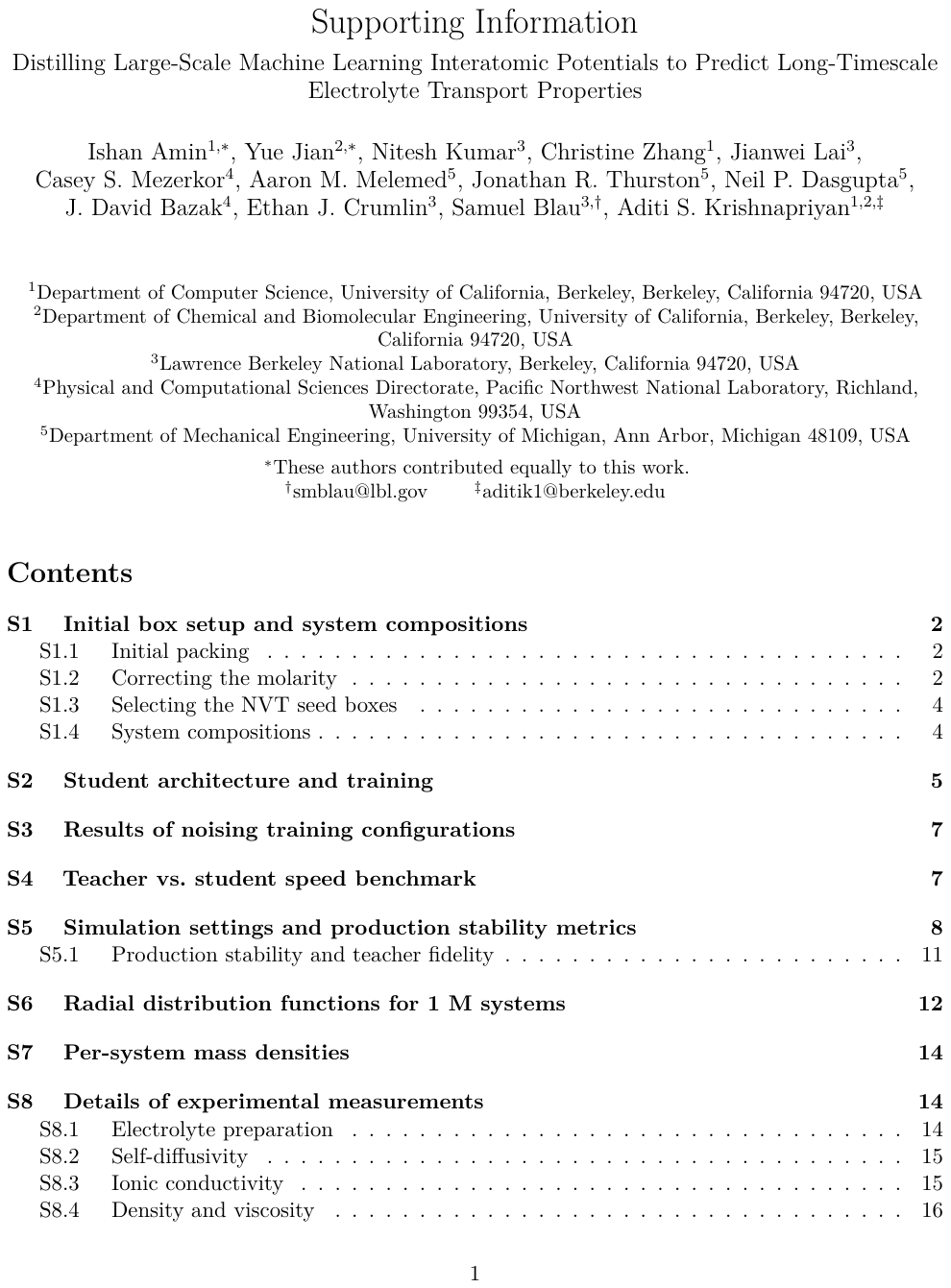}\edef\SIpages{\the\pdflastximagepages}%
  \else
    \edef\SIpages{\the\XeTeXpdfpagecount"appendix.pdf" }%
  \fi
  \foreach \p in {1,...,\SIpages}{%
    \clearpage\thispagestyle{empty}\noindent
    \makebox[0pt][l]{\hspace*{-\dimexpr1in+\hoffset+\oddsidemargin\relax}%
      \raisebox{\dimexpr1in+\voffset+\topmargin+\headheight+\headsep+\topskip-\height\relax}[0pt][0pt]{%
        \includegraphics[page=\p,width=\paperwidth,height=\paperheight]{appendix.pdf}}}%
  }%
}{}
\end{document}